\documentclass[aps,pra,twocolumn,showpacs,floatfix]{revtex4} 
\usepackage{amsmath,amssymb,isolatin1,graphicx,times}

\newcommand{\be}{\begin{equation}}
\newcommand{\ee}{\end{equation}}
\newcommand{\bea}{\begin{eqnarray}}
\newcommand{\eea}{\end{eqnarray}}
\newcommand{\bw}{\begin{widetext}}
\newcommand{\ew}{\end{widetext}}

\newcommand{\kommentar}[1]{}

\begin{document}
 
\title{Localization of coherent exciton transport in phase space 
}
\author{Oliver M{\"u}lken}
\author{Veronika Bierbaum}
\author{Alexander Blumen}
\affiliation{
Theoretische Polymerphysik, Universit\"at Freiburg,
Hermann-Herder-Straße 3, D-79104 Freiburg, Germany}

\date{\today} 
\begin{abstract}
We study numerically the dynamics of excitons on discrete rings in the
presence of static disorder. Based on continuous-time quantum walks we
compute the time evolution of the Wigner function (WF) both for pure
diagonal (site) disorder, as well as for diagonal and off-diagonal (site
and transfer) disorder. In both cases, large disorder leads to
localization and destroys the characteristic phase space patterns of the
WF found in the absence of disorder. 
\end{abstract}
\maketitle

\section{Introduction}

Ever since the emergence and development of quantum mechanics, there has
been a major interest in the crossover from quantum mechanical transport
to the corresponding classical transport.  However, there are far-reaching
differences in the usual mathematical description of the two different
processes. In classical physics, the phase space is spanned by
conjugate variables, such as position and momentum, whose time development
leads to classical transport. Quantum mechanical processes,
on the contrary, take place in Hilbert space. One approach to overcome
these differences was already presented in the early days of quantum
mechanics, more than 70 years ago, by Wigner \cite{Wigner1932,
Hillery1984}. He introduced a function, now known as Wigner function (WF),
which is a (quasi) probability in a quantum mechanical phase space spanned
by position and momentum variables. The WF is a real valued function and
in this respect compares well with the classical Boltzmann probability
distribution in phase space.  However, it is not always positive. WFs and
related phase space functions, like the Husimi function, are widely used
in Quantum Optics \cite{Schleich,Mandel-Wolf} but also for describing
electronic transport, see e.g.\ \cite{kluksdahl1989,buot1993,bordone1999}.

Apart from the quantum-classical crossover there are also extremely
interesting purely quantum mechanical phenomena. For example, Anderson has
shown that there is no quantum diffusion for some random lattices
\cite{anderson1958,anderson1978,abrahams1979}, an effect nowadays called
(strong) localization, where the quantum mechanical transport through the
lattice is, in essence, prohibited by the potential energy surface.
Anderson's hopping model for electron transport has also turned out to be
useful in describing excitons in disordered systems, e.g.,
\cite{logan1987,abramavicius2004,dominguez2004,helmes2005,barford2006}.

In fact, Anderson's model is closely related to the
so-called continuous-time quantum walks (CTQWs) with disorder.  Recently,
it has been shown that the motion on a graph, described by CTQWs, can be
exponentially suppressed by the disorder \cite{keating2006}.
Originally, (unperturbed) CTQWs were introduced in the context of quantum
information as the quantum mechanical analog of continuous-time random
walks (CTRWs) \cite{farhi1998}. Here, the underlying, discrete
connectivity of the structure on which the transport takes place
determines the Hamiltonian. Since CTQWs also model exciton
transport over various discrete structures \cite{mbb2006a},
they are closely connected to other appraches to study
(coherent) exciton transport phenomena on discrete graphs, for instance,
in the contexts of polymer \cite{Kenkre}, atomic \cite{Shore} or
solid-state \cite{Haken} physics. 

In this paper we consider the dynamics of excitons on discrete
rings under static disorder, focussing on a quantum mechanical phase space
approach.  We compute the corresponding WFs numerically and study the
effects of both pure diagonal (site) disorder, as well as of diagonal and
off-diagonal (site and transfer) disorder on the (coherent) transport. The
WFs are then compared to the unperturbed case, for which an analytical
treatment is possible \cite{mb2006a}.

The paper is organized as follows. In Sec.~\ref{ced} we briefly review how
we model coherent exciton transport on graphs and define the appropriate
Hamiltonian for rings with two types of disorder. After introducing the
discrete WF in Sec.~\ref{wf}, we lay down the procedure of our
calculations in Sec.~\ref{wfdis}. The subsequent Secs.~\ref{wfdis_result}
to \ref{wf_lt_result} show the numerical results for coherent exciton
transport on rings with disorder. We close with our conclusions.

\section{Coherent exciton dynamics}\label{ced}

In the absence of disorder, the (coherent) dynamics of excitons on a graph
of connected nodes is modelled by the CTQW. Here, the CTQW is obtained by
identifying the Hamiltonian of the system with the (classical) transfer
matrix, ${\bf H} = - {\bf T}$, see e.g.\
\cite{farhi1998} (we will set $\hbar \equiv 1$ in the following). The
transfer matrix of the walk, ${\bf T} = (T_{l,j})$, can be related to
the connectivity matrix ${\bf A}$ of the graph by ${\bf T} = - \gamma {\bf
A}$, where for simplicity we assume the transmission rates $\gamma$ of all
bonds to be equal.  The matrix ${\bf A}$ has as non-diagonal elements
$A_{l,j}$ the values $-1$ if nodes $l$ and $j$ of the graph are connected
by a bond and $0$ otherwise. The diagonal elements $A_{j,j}$ of ${\bf A}$
equal the number of bonds $f_j$ which exit from node $j$. 

Now, the states $|j\rangle$ associated with excitations localized at the
nodes $j$ span the whole accessible Hilbert space to be considered here.
In general, the time evolution of a state $|j\rangle$ starting at time
$t_0=0$ is given by $| j;t \rangle = \exp(-i {\bf H} t)|j\rangle$. By
denoting the eigenstates of ${\bf H}$ by $|\Phi_\theta\rangle$ and the
eigenvalues by $E_\theta$ the transition probability reads $\pi_{l,j}(t)
\equiv \left|\langle l | \exp(-i {\bf H} t) | j \rangle\right|^2 = \left|
\sum_\theta \exp(-iE_\theta t) \langle l | \Phi_\theta \rangle \langle
\Phi_\theta | j\rangle \right|^2$.

\subsection{Dynamics on rings without disorder}\label{ctqw_ring_0}

The unperturbed Hamiltonian for such a CTQW on a finite one-dimensional
network of length $N$ with periodic boundary conditions (PBC) takes on the
very simple form ${\bf H}^{0} |j\rangle = 2 |j\rangle - |j-1\rangle -
|j+1\rangle$, where we have taken the transmission rate to be
$\gamma\equiv 1$ and $j=0,1,\dots,N-1$. The eigenstates
$|\Phi_\theta^{0}\rangle$ of the time independent Schr\"odinger equation
${\bf H}^{0} |\Phi_\theta^{0}\rangle = E_\theta^{0}
|\Phi_\theta^{0}\rangle$ are Bloch states, which can be expressed as
linear combinations of the states $|j\rangle$, see Refs.\
\cite{mb2005b,mb2006a} for details.

\subsection{Dynamics on rings with disorder}\label{ced_d}

Now we introduce (static) disorder by adding to the unperturbed
Hamiltonian ${\bf H}^{0}$ a disorder operator $\mathbf \Delta$, i.e., by
setting ${\bf H} = {\bf H}^{0} + {\mathbf \Delta}$. We let the disorder
matrix $\mathbf \Delta = (\Delta_{l,j})$ have non-zero entries only at the
positions for which $H_{l,j}\neq0$. For different strengths of disorder,
the elements $\Delta_{l,j} = \Delta_{j,l}$ are chosen randomly (drawn
from a normal distribution with the mean value zero and the variance one),
which we multiply by a factor of $\Delta$, which then can take values from
the interval $[0,1/2]$.  Note that under these assumptions for the
(static) disorder the connectivity of the graph is essentially unchanged,
i.e., there are no new connections created nor are existing connections
destroyed. Therefore, the only non-zero matrix elements of ${\bf H}$ are
those of the initial ${\bf A}$.
The action of the new Hamiltonian ${\bf H}$ on a state $|j\rangle$ reads
thus
\bea
{\bf H} |j\rangle &=& \Big({\bf H}^{0} + {\mathbf \Delta}\Big) |j\rangle =
2 |j\rangle - |j-1\rangle - |j+1\rangle \nonumber \\
&+& 2\Delta_{j,j} |j\rangle - \Delta_{j,j-1} |j-1\rangle - \Delta_{j,j+1}
|j+1\rangle.
\eea 

In the following we consider two cases of disorder: 

(A) 
Diagonal disorder (DD), where $\Delta_{j,j}\ne0$ and $\Delta_{l,j}=0$ for
$l\ne j$.  We assign a random number to each
$\Delta_{j,j}$, a procedure which leads to $N$ random numbers.

(B) 
Diagonal and off-diagonal disorder (DOD), where we choose a random number
for each $\Delta_{j,j}$ and for each $\Delta_{j,j-1}$. Thus, $2N$ random
numbers are needed.

Introducing disorder into the system in this way has consequences for the
relation between the CTQWs and the CTRWs. In CTRWs the transition rates,
given by the entries of the the transfer matrix ${\bf T}$, are correlated,
i.e., for each site the sum of the non-diagonal rates for transmission
from it and the diagonal rate of leaving it are the same.  In the cases
considered here, a direct identification of the Hamiltonian ${\bf H}$ with
a classical transfer matrix ${\bf T}$ is, in general, not possible
anymore. However, the DOD and DD Hamiltonians are widely used in quantum
mechanical nearest-neighbor hopping models, to which also the CTQWs
belong.  Furthermore, we still consider transport processes on graphs
which have the connectivity matrix ${\bf A}$, but the direct connection
between ${\bf H}$ and ${\bf T}$ is lost.  Of course, one can maintain this
connection by imposing constraints on the $\Delta$, e.g., by requiring
that $2\Delta_{j,j} = \Delta_{j,j-1} + \Delta_{j,j+1}$ for all $j$.

\section{Wigner functions}\label{wf}

The WF is a quasi-probability (in the sense that it can become negative)
in the quantum mechanical phase space. For a phase space spanned by the
continuous variables $X$ and $K$, the WF reads
\cite{Wigner1932,Hillery1984,Schleich}
\be
W(X,K;t) = \frac{1}{\pi}\int \ dY \ e^{iKY} \ \langle X - Y/2 | \hat
\rho(t) | X + Y/2
\rangle ,
\label{wigner}
\ee
where $\hat \rho(t)$ is the density operator. In the following we only
consider pure states, i.e., $\hat \rho (t) = |j;t\rangle \langle j;t|$.
Note that the WF is normalized to one when integrated over the whole phase
space.

For a discrete system having $N$ sites on a ring, where we choose to
enumerate the sites as $0,1,\dots,N-1$, $\langle x | j;t\rangle$ is only
defined for integer values of $x=0,1,\dots,N-1$, and the form of
Eq.(\ref{wigner}) has to be changed from an integral to a sum.  Now the WF
resembles a discrete Fourier transform, which requires $N$ different
$k$-values. These $k$-values may evidently be chosen as $k=2\pi\kappa/N$,
again having $\kappa=0,1,\dots,N-1$. According to \cite{mb2006a} we use,
for integer $x$ and $y$, the following discrete WF
\be
W_j(x,\kappa;t) \equiv \frac{1}{N} \sum_{y=0}^{N-1} e^{iky} \ \langle x-y | j;t
\rangle \langle j;t | x+y \rangle .
\label{wigner_discrete}
\ee
The marginal distributions are now given by summing over lines in phase
space, e.g., we get 
\bea
\sum_{\kappa=0}^{N-1} W_j(x,\kappa;t) &=& \frac{1}{N}
\sum_{\kappa=0}^{N-1} \sum_{y=0}^{N-1} e^{i2\pi\kappa y/N} \ \langle x-y |
j;t
\rangle \langle j;t | x+y \rangle  \nonumber \\
&=& \sum_{y=0}^{N-1} \delta_{y,o} \ \langle x-y |
j;t
\rangle \langle j;t | x+y \rangle \nonumber \\
&=& |\langle x | j;t \rangle|^2 \equiv \pi_{x,j}(t).  
\label{wigner_discrete_marg}
\eea
We note here that
there are forms similar to Eq.~(\ref{wigner_discrete}) in defining
discrete WFs, see, e.g., Refs.\
\cite{Wootters1987,Cohendet1988,Leonhardt1995,Takami2001,Miquel2002}.

In order to compare to the classical long time behavior, we define a long
time average of the WF by
\be
{\overline W}_j(x,\kappa)
\equiv \lim_{T\to\infty}\frac{1}{T} \int\limits_0^T dt \
W_j(x,\kappa;t).
\label{wf_tavg}
\ee
As for the marginal distributions for the WF we obtain marginal
distributions for ${\overline W}_j(x,\kappa)$ upon integration along lines
in $(x,\kappa)$-space. Especially, summing over $\kappa$, $\sum_{\kappa}
{\overline W}_j(x,\kappa) \equiv \chi_{xj}$, one obtains the
long time average of the transition probability $\pi_{x,j}(t)$.

\section{WFs on rings}\label{wfdis}

\subsection{WFs on rings without disorder}

The unperturbed Hamiltonian for a CTQW on a ring is given by ${\bf
H}^{0}$, see Sec.~\ref{ctqw_ring_0}. In this case, the discrete WF can
be calculated analytically by using a Bloch ansatz. It is given by
\cite{mb2006a}
\bea
&& W_j(x,\kappa;t) = \frac{1}{N^2} \sum_{n=0}^{N-1} \exp\big[
-i2\pi(2n+\kappa)(x-j)/N \big] \nonumber \\
&& \times \exp\big\{ -i2t[\cos(2\pi (\kappa + n)/N) - \cos(2\pi n/N)]
\big\} .
\label{wigner_bloch_3}
\eea
As shown in Ref.\ \cite{mb2006a}, the marginal distributions of
$W_j(x,\kappa;t)$ and ${\overline W}_j(x,\kappa)$ are correctly recovered. 

\subsection{WFs on rings with disorder}

Under disorder the Bloch property is lost, then the previous analytic
approach does not apply anymore.  We hence compute numerically the WFs for
the corresponding quantum dynamics of excitons on rings with disorder by
using the standard software package MATLAB. The differences in the
following computations are only due to the different Hamiltonians ${\bf
H}$, depending on the specific type of disorder.

Having the eigenstates $|\Phi_{\theta}\rangle$ and the eigenvalues
$E_\theta$, the WF is obtained by expanding Eq.~(\ref{wigner_discrete}) as
\bea
W_j(x,\kappa;t) &=& \frac{1}{N} \sum_{y=0}^{N-1} e^{iky}
\sum_{\theta,\theta'} \exp[-i(E_{\theta'}-E_\theta)t] 
\nonumber \\
&\times& \langle x-y | \Phi_{\theta'} \rangle \langle \Phi_{\theta'} | j\rangle \langle j |
\Phi_\theta
\rangle \langle \Phi_\theta | x+y
\rangle
.
\label{wf_noise}
\eea

\begin{figure}[htb]
\centerline{\includegraphics[clip=,width=0.99\columnwidth]{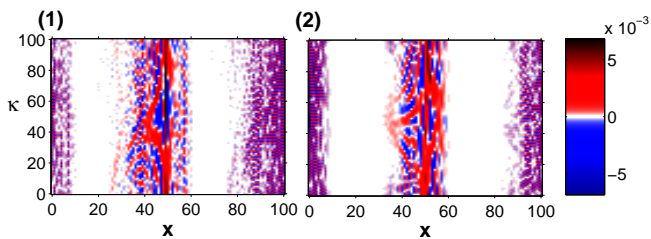}}
\caption{(Color online)
Snapshots of two realizations of the WF for $N=101$ and DOD with $\Delta=1/2$
at time $t=40$.
}
\label{wigner_101_one}
\end{figure}

Figure \ref{wigner_101_one} shows the WF according to Eq.(\ref{wf_noise})
for $N=101$ and two different realizations of ${\bf H}$ with $\Delta=1/2$
at time $t=40$. Clearly, the details of the WF differ much for
different realizations. However, a trend of the effect of disorder on the
dynamics is already visible: The disorder prevents the excitation to
travel freely through the graph. Instead, there is a localized area about
the initial site $x=j=50$. We will study this in much more detail in the
remainder.

There appear non-vanishing values of the WF at nodes opposite to the
initial node $j$, i.e., at $x \approx j+N/2$, even at infinitesimal small
times $t$.  These values are not related to disorder but are due to the
PBCs. This can be inferred from Eq.~(\ref{wigner_discrete}). Consider
first the case of even $N$: for $t=0$ and $x=j$ the only contributions to
the sum in Eq.~(\ref{wigner_discrete}) are those for which $y=0$. However,
for $t=0$ and also for $x=j+N/2$ the WF does not vanish: There are
non-zero contributions to the sum for $y=N/2$, since we have $ \langle j |
j \rangle \langle j | j + N \rangle = \langle j | j \rangle\langle j | j
\rangle = 1$.  Non-zero values of the WF for $x=j+N/2$ continue to show up
at later times, see also \cite{mb2006a}. The argumentation for odd $N$ is
similar: Here, however, non-vanishing values at $x\approx j+N/2$ start to
appear as soon as the wave function $\langle x | j;t\rangle$ spreads over
more than one node. The difference between these non-vanishing WF-values
for even and for odd $N$ at short times tends to zero as $1/N$ for $N$
large.  We remark that the appearence of these non-vanishing WF-values for
short times at $x \approx j+N/2$ does not lead to a finite transition
probability $\pi_{j+N/2,j}(t)$, since at small $t$ the sum in
Eq.(\ref{wigner_discrete_marg}) practically vanishes. This is in line with
the situation for open boundaries, where the transition probabilities
$\pi_{j+N/2,j}(t)$ also vanish at short times. There, however, also the
WF-values at $x \approx j+N/2$ are practically zero at short times.  A
thorough study of the influence of different boundary conditions on the WF
is beyond the scope of this paper and will be given elsewhere.

Once we have the WF, we also compute its long time average.  Now it is
preferrable not to first evaluate Eq.~(\ref{wf_noise}) and to perform then
the computationally expensive time integrals of Eq.~(\ref{wf_tavg}), but
to proceed at first analytically, obtaining
\bea
{\overline W}_j(x,\kappa) 
&=& \frac{1}{N} \sum_y e^{iky} \sum_{\theta,\theta'} \tilde\delta(E_\theta - E_{\theta'})
\nonumber \\
&\times& \langle x-y | \Phi_{\theta'} \rangle \langle \Phi_{\theta'}
|j \rangle \langle j | \Phi_\theta \rangle \langle \Phi_\theta
|x+y \rangle ,
\label{wf_tavg_2}
\eea
with $\tilde\delta(E_\theta - E_{\theta'}) = 1$ if $E_\theta = E_{\theta'}$ and
$\tilde\delta(E_\theta - E_{\theta'}) = 0$ otherwise. Eq.~(\ref{wf_tavg_2}) is
in general much more accurate and computationally cheaper.

\subsection{Ensemble averages}

In order to have a global picture of the effect of disorder on the
dynamics, we will consider ensemble averages of the WFs. For this we
calculate the WF for different realizations of ${\bf H}$ and average over
all realizations, i.e., for $R$ realizations we compute
\be
\langle W_j(x,\kappa;t) \rangle_R \equiv \frac{1}{R} \sum_{r=1}^R
\big[W_j(x,\kappa;t)\big]_r,
\ee
where $\big[W_j(x,\kappa;t)\big]_r$ is the WF of the $r$th realization of
${\bf H}$.

\section{The role of disorder}\label{wfdis_result}

For all cases of disorder, we study the WF for different strength of the
disorder, i.e., for different $\Delta$ with $\Delta$ ranging from
$\Delta=1/40$ to $\Delta=1/2$. We exemplify our results for odd ($N=101$)
and even ($N=100$) numbered rings, where in all cases the initial
condition ($t=0$) is a state localized at node $j$ (in our two examples
$j=50$). Note that the WF of a localized state at $x=j$ is equipartitioned
in the $\kappa$-direction, i.e., $W_j(j,\kappa;0)=1/N$.

\subsection{Diagonal disorder (DD)}

\begin{figure}[htb]
\centerline{\includegraphics[clip=,width=0.99\columnwidth]{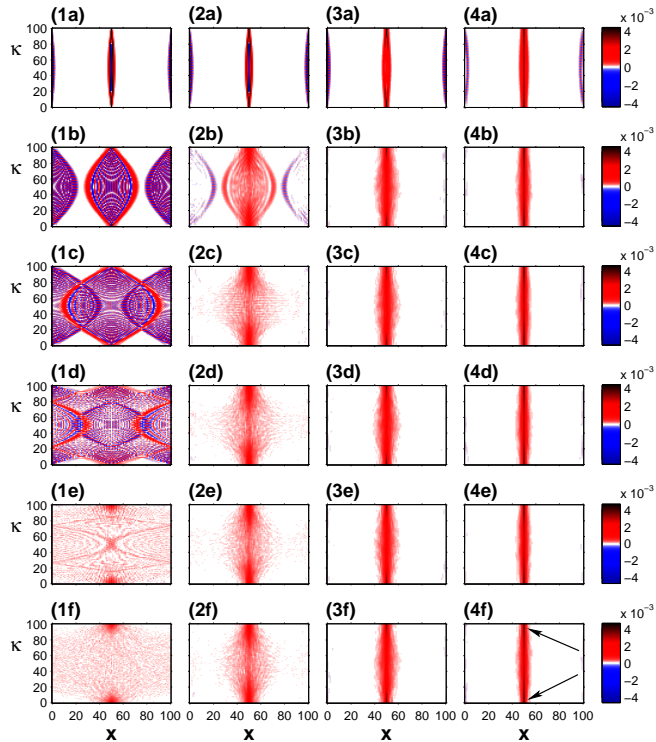}}
\caption{(Color online) 
DD:
Ensemble average of WFs of the quantum dynamics on a ring of length $N=101$ for
$\Delta=1/40,1/10,1/4$, and $1/2$ [columns (1)-(4)], each at times
$t=1,10,20,40,100$, and $500$ [rows (a)-(f)]. The initial node is always
$j=50$ and the average is over $R=1000$ realizations. Red regions denote
positive values of the averaged WFs, blue regions negative values and
white regions values close to $0$. The colormaps are always chosen to be
the same for each row but might differ in different rows.
The maximal values of $\langle W_j(x,\kappa;t) \rangle_R$ are denoted by
small black regions; these are highlighted and   exemplified by the arrows
in panel (4f).
}
\label{wigner_101_diag}
\end{figure}

We start by considering diagonal disorder for a graph with $N=101$.
Figure~\ref{wigner_101_diag} shows snapshots of $\langle W_j(x,\kappa;t)
\rangle_R$ for different values of $\Delta$ at different times. 

The first column shows $\langle W_j(x,\kappa;t) \rangle_R$ for
$\Delta=1/40$ at times $t=1,10,20,40,100$, and $500$
[Fig.~\ref{wigner_101_diag}(1a)-(1f)].  For this quite weak disorder, the
patterns in phase space are similar to the unperturbed case, where
``waves'' in phase space emanate from the initial site $x=j=50$ and start
to interfere after having reached the opposite site of the ring (see
Fig.~3 of \cite{mb2006a}). However, at longer times differences become
visible, Fig~\ref{wigner_101_diag}(1f).  The pattern for the unperturbed
case is quite irregular but with alternating positive and negative regions
of the WF of approximately the same magnitude. Here, the disorder causes a
decrease of $\langle W_j(x,\kappa;t) \rangle_R$ for $x$ close to the
initial site $j=50$ and $\kappa$-values in the middle of the interval
$[0,N-1]$ compared to the values of $\kappa$ close to $0$ or $N-1$.

Increasing the disorder parameter $\Delta$, the patterns change
profoundly. The wave structure gets suppressed and for all $\kappa$ a
localized region forms about the initial site $j=50$ already for small
disorder ($\Delta=1/10$) and short times ($t=20$), see
Fig.~\ref{wigner_101_diag}(2c).

For even larger values of $\Delta$, for all $\kappa$ the formation of a
localized region about $j=50$ becomes even more pronounced. Already for
$\Delta=1/4$ this localized region forms for times as short as $t=10$, see
Fig.~\ref{wigner_101_diag}(3b).  At $\Delta=1/2$, $\langle W_j(x,\kappa;t)
\rangle_R$ stays localized for all times
[Fig.~\ref{wigner_101_diag}(4a)-(4f)].  Also here, values of the WF at
about $\kappa\approx N/2$ are decreased, whereas values of the WF for
$\kappa$ about the interval borders $0$ and $N-1$ remain rather large for
$x\approx j$, as indicated by the thin black region (see also the arrows),
e.g., in Fig.~\ref{wigner_101_diag}(4f).  We further note that at high
disorder the localized averaged WF is always positive, i.e., all
fluctuations, present for small disorder, have vanished. We recall that
the WF is normalized to one when integrated over the whole phase space.

\begin{figure}[htb]
\centerline{\includegraphics[clip=,width=0.99\columnwidth]{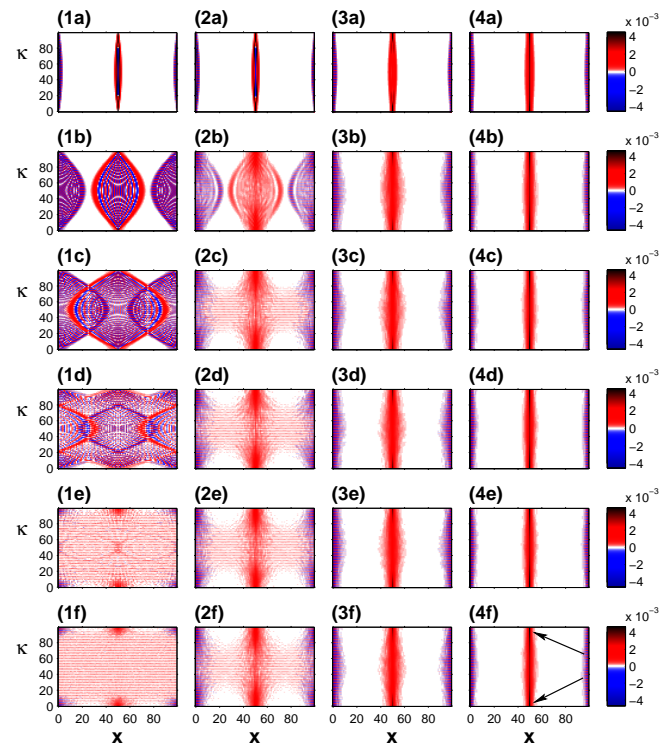}}
\caption{(Color online) 
DD:
Same as Fig.~\ref{wigner_101_diag} but for $N=100$.}
\label{wigner_100_diag}
\end{figure}

Having an even number of nodes in the graph does not alter the picture
drastically. By comparing Fig.~\ref{wigner_101_diag} to
Fig.~\ref{wigner_100_diag}, which shows the quantum dynamics on a ring of
$N=100$ nodes with DD, one sees that the corresponding panels are rather
similar; the localization effect of the disorder on the system stays the
same. However, there are also differences at long times, compare (1e-f)
and (2e-f) in each figure. Another difference between even and odd $N$
lies in the fact that for even $N$ there remains a region of alternating
values of $\langle W_j(x,\kappa;t) \rangle_R$ at about $x=j+N/2$, even for
large disorder, see column (4a)-(4f).  This is due to the periodic
boundary condition and, therefore, to the equal number of steps in both
directions starting from any site on the network needed to reach the
opposite site of the network. Obviously, the disorder does not destroy
this symmetry.

\subsection{Diagonal and off-diagonal disorder (DOD)}

\begin{figure}[htb]
\centerline{\includegraphics[clip=,width=0.99\columnwidth]{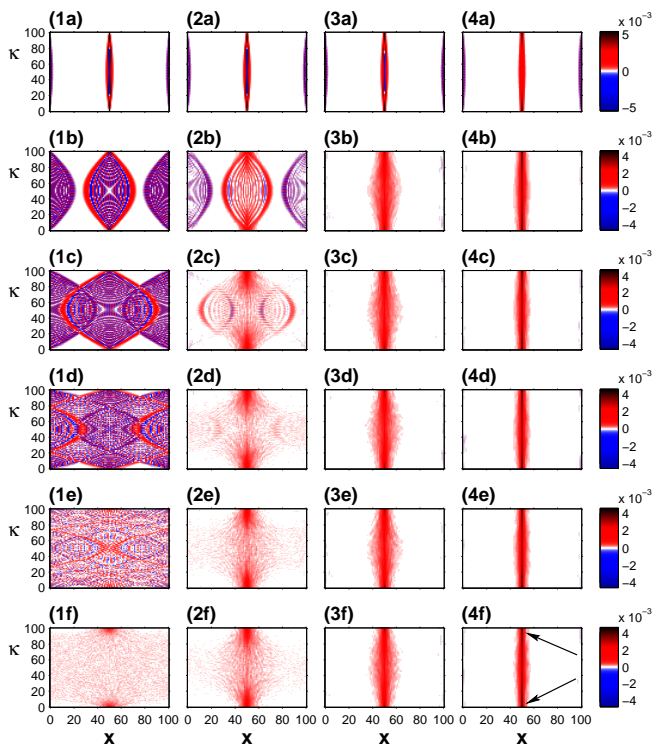}}
\caption{(Color online) 
DOD:
Same values of $t$, $\Delta$, and $R$ as used in Fig.~\ref{wigner_101_diag}.}
\label{wigner_101_siteoff}
\end{figure}

The second type of disorder is the diagonal and off-diagonal one (DOD),
where both the diagonal and the off-diagonal elements of the unperturbed
Hamiltonian are randomly changed, as explained in Sec.~\ref{ced_d}.
Figure~\ref{wigner_101_siteoff} shows $\langle W_j(x,\kappa;t) \rangle_R$
for the same values of $t$, $\Delta$, and $R$ as in
Fig.~\ref{wigner_101_diag}.  Here and as for DD, the symmetry with respect
to the line in phase space at $x=j=50$ remains intact. Comparing to the
DD case, there is a slightly weaker suppression of the waves in phase
space.  This effect is rather small but can be seen by comparing, e.g.,
the values of the WF in Fig.~\ref{wigner_101_diag}(2f) to the ones in
Fig.~\ref{wigner_101_siteoff}(2f), note the different corresponding
colorbars. However, the overall localization effect is very similar for
both types of disorder.

\begin{figure}[htb]
\centerline{\includegraphics[clip=,width=0.99\columnwidth]{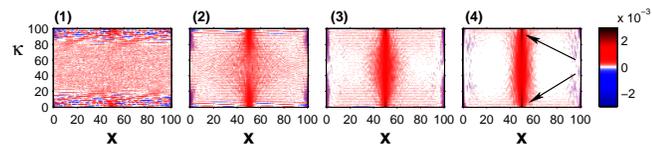}}
\caption{(Color online)
Constrained diagonal and off-diagonal disorder, see text for details:
Averaged WF for $N=101$ and $t=500$ [cases (f) in
Figs.~\ref{wigner_101_diag} to \ref{wigner_101_siteoff}], but for the same
values of $\Delta$ and $R$ as used in Fig.~\ref{wigner_101_diag}.}
\label{wigner_101_korr}
\end{figure}

Another possibility is to constrain the disorder by having $2\Delta_{j,j}
= \Delta_{j,j-1} + \Delta_{j,j+1}$, which maintains the connection of
${\bf H}$ to the classical transfer matrix ${\bf T}$.
Figure~\ref{wigner_101_korr} shows the WF evaluated in this way for
$N=101$ at time $t=500$ and for $\Delta= 1/40,~1/10,~1/4$ and $1/2$. Also
here, the localization effect is clearly seen. Nevertheless, the details
of the WF differ from those obtained for the other types of disorder.

\section{Marginal distributions}

\begin{figure}[htb]
\centerline{\includegraphics[clip=,width=0.99\columnwidth]{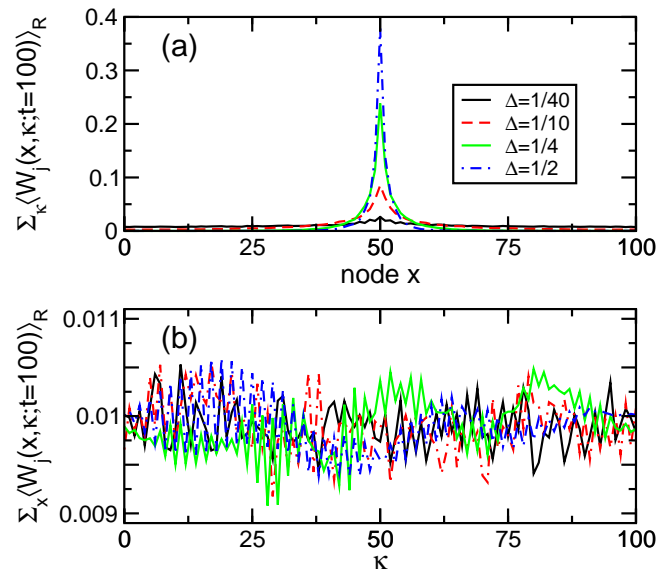}}
\caption{(Color online)  
DOD:
Marginal distributions of $\left<W_j(x,\kappa;t) \right>_R$, for
$N=101$ at $t=100$, for $\Delta=1/40,1/10,1/4$, and $1/2$. (a) summing
along the $\kappa$ direction, (b) summing along the $x$ direction.}
\label{wigner_marg}
\end{figure}

Integrating the WF along lines in phase space gives the marginal
distributions. Since we saw that the effect of localization does not
depend on the particular choice of the type of disorder, we display the
marginal distributions only for DOD. Figure \ref{wigner_marg} shows the
marginal distributions for $N=101$ at time $t=100$ for different $\Delta$.
Obviously, the details of the WF are lost when integrating along either
the $x$ or the $\kappa$ direction. 

While the transition probabilities, obtained by summing over all $\kappa$
[Fig.\ref{wigner_marg}(a)], clearly show the effect of localization, this
is not the case for the marginal distribution obtained by summing over all
$x$ [Fig.\ref{wigner_marg}(b)]. In the latter case, one cannot readily
distinguish the situations corresponding to different values of $\Delta$.

\section{Long time averages}\label{wf_lt_result}

In has been shown in \cite{mb2006a} that for unperturbed coherent exciton
transport and odd $N$ (superscript $^o$), most points in the quantum
mechanical phase space have a weight of $1/N^2$, namely we have 
\be
{\overline W}^o_j(x,\kappa) = \begin{cases} 1/N^2 & \kappa\neq0 \
\mbox{and any} \ x \\ 1/N & \kappa=0 \ \mbox{and} \ x=j \\ 0 &
\mbox{else.} \end{cases}
\label{lim_wf_0}
\ee
For even $N$, the limiting WF reads
\be
{\overline W}^e_j(x,\kappa) = \begin{cases} 2/N^2 & \kappa\neq0, \ \kappa
\ \mbox{even}\ \mbox{and any} \ x \\ 1/N & \kappa=0 \ \mbox{and} \
x=j,j+ N/2 \\ 0 & \mbox{else.} \end{cases}
\label{lim_wf_0e}
\ee
We note that Eq.~(\ref{lim_wf_0e}) differs from Eq.~(19) of
Ref.~\cite{mb2006a}, which is not correct. The long time averages of the
WFs for even $N$ are somewhat peculiar, since values different from zero
appear only for even $\kappa$, whereas the WFs themselves have values
different from zero at arbitrary times for all $\kappa$. A numerical
check (which we do not include here) for a finite line of $N$ nodes without
disorder, shows that these stripes in the long time average in the present
study are due to the periodic boundaries. For even $N$ there are
constructive interference patterns in the transition probabilities, since
the number of steps in both directions is the same, see also
Ref.~\cite{mb2005b}. For a finite line with even $N$, there are no stripes
in the long time average. Of course, for short times and close to the
initial node, the WFs for the line and the circle coincide,
since the boundaries have no effect on the initial propagation.
Nevertheless, one also has to bear in mind that the long time average is
not a real equilibrium distribution.

As a technical note, we remark that changing the order of the time and
ensemble averages can lead to a considerable speed-up of the numerical
computation. We checked numerically that the time average and the ensemble
average indeed interchange, i.e., we have
\bea
\langle {\overline W}_j(x,\kappa)\rangle_R
&\equiv& \Big\langle\lim_{T\to\infty}\frac{1}{T} \int\limits_0^T dt \
W_j(x,\kappa;t) \Big\rangle_R \nonumber \\
&=& \lim_{T\to\infty}\frac{1}{T} \int\limits_0^T dt \
\langle W_j(x,\kappa;t) \rangle_R.
\label{wf_avg_tavg}
\eea
Now, using Eq.~(\ref{wf_tavg_2}) reduces the computational effort further.
For computational reasons, we further interchanged the summation over
$y$ with the ensemble average, i.e.,
\bea
&&\langle{\overline W}_j(x,\kappa)\rangle_R
= \Big\langle\frac{1}{N} \sum_y e^{iky} \sum_{\theta,\theta'} \tilde\delta(E_\theta -
E_{\theta'}) 
\nonumber \\
&&\qquad \times \langle x-y | \Phi_{\theta'} \rangle \langle \Phi_{\theta'}
|j \rangle \langle j | \Phi_\theta \rangle \langle \Phi_\theta
|x+y \rangle\Big\rangle_R \nonumber \\
&&= \frac{1}{N} \sum_y e^{iky} \Big\langle\sum_{\theta,\theta'} \tilde\delta(E_\theta -
E_{\theta'}) 
\nonumber \\
&&\qquad \times \langle x-y | \Phi_{\theta'} \rangle \langle \Phi_{\theta'}
|j \rangle \langle j | \Phi_\theta \rangle \langle \Phi_\theta
|x+y \rangle\Big\rangle_R.
\label{wf_avg_tavg_2}
\eea
Again, we carefully checked that Eq.~(\ref{wf_avg_tavg_2}) yields the same
result as Eq.~(\ref{wf_avg_tavg}). 

\begin{figure}[htb]
\centerline{\includegraphics[clip=,width=0.99\columnwidth]{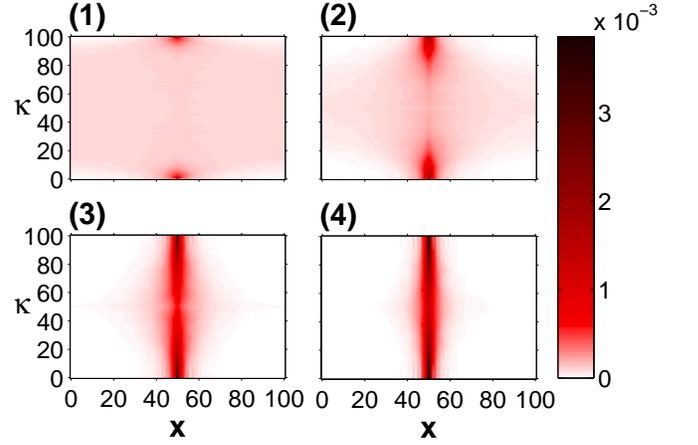}}
\caption{(Color online)  
DOD: Limiting averaged WF, $\langle{\overline W}_j(x,\kappa)\rangle_R$, for
$N=101$ and $\Delta=1/40$, $1/10$, $1/4$, and $1/2$ [panels (1) to (4),
respectively], according to Eq.(\ref{wf_avg_tavg_2}).}
\label{lim_wigner}
\end{figure}

Now, the disorder changes also the limiting WF quite drastically.
Starting from high disorder of $\Delta=1/2$, we expect from
Figs.~\ref{wigner_101_diag}(4a)-(4f) to \ref{wigner_101_siteoff}(4a)-(4f)
that the long time average of the {\sl averaged} WF will look basically
the same. Figure \ref{lim_wigner} shows the limiting averaged WF for
$N=101$ and DOD according to Eq.~(\ref{wf_avg_tavg_2}). Indeed, for large
$\Delta$, the limiting averaged WF is comparable to the corresponding
averaged WF, compare Figs.~\ref{lim_wigner}(4) and
\ref{wigner_101_siteoff}(4f). Close to the initial node $x=j=50$ and there
along the $\kappa$-direction, $\langle{\overline W}_j(x,\kappa)\rangle_R$
has large (positive) values for $\kappa$ about $0$ and $N-1$ which
decrease by going toward $\kappa=N/2$. Here, the minimum of this decrease
depends on the particular type of disorder.

Also for small $\Delta$ there are significant differences to the case
without disorder. From Fig.~\ref{lim_wigner}(1) we see that the disorder
``smears out'' the localized value $\overline{W}_j(j,0)=1/N$ in the case
without disorder, see Eq.(\ref{lim_wf_0}). Specifically, the
$(x=j,\kappa=0)$-value decreases while the neighboring ones increase.  For
$\Delta=1/10$, the onset of localization about $x=j=50$ is already seen, see
Fig.~\ref{lim_wigner}(2), and becomes more and more pronounced as $\Delta$
increases, Fig.~\ref{lim_wigner}(3). Furthermore, all other values of
$\langle{\overline W}_j(x,\kappa)\rangle_R$ for $x\neq j$ decrease with
increasing disorder, as can be seen by the decreasing size of the light
pink region, which corresponds to values close to $1/N^2$.  In fact, the
values of the limiting averaged WF for $x$ distant from $x=j=50$ drop to
zero, shown as white regions in Fig.~\ref{lim_wigner}.

\begin{figure}[ht]
\centerline{\includegraphics[clip=,width=0.99\columnwidth]{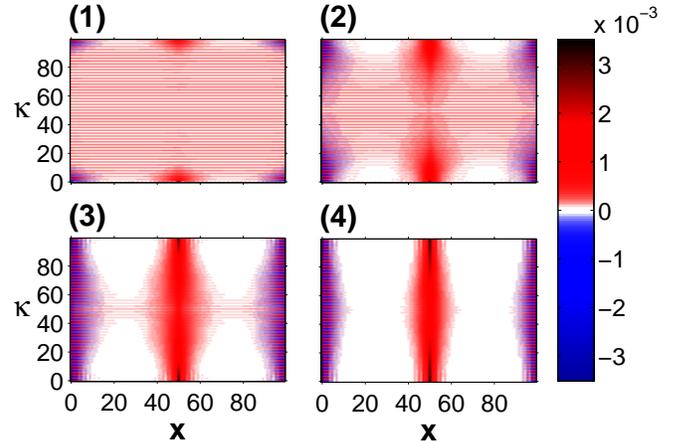}}
\caption{(Color online)  
Same as Fig.~\ref{lim_wigner} but for $N=100$.}
\label{lim_wigner_even}
\end{figure}

For even $N$ we found that without disorder the limiting WF shows a
peculiar ``striped'' distribution, caused by the PBCs, see
Eq.(\ref{lim_wf_0e}). By switching
on the disorder, these peculiarities vanish. Figure \ref{lim_wigner_even}
shows the limiting averaged WF for DOD and $N=100$.  Although for small
$\Delta$ there are some remainders of stripes left, these disappear
completely for higher values of $\Delta$, compare
Figs.~\ref{lim_wigner_even}(1)-(4). Note further that the second peak of
${\overline W}^e_j(x,\kappa)$ at $x=j+N/2$, see Eq.(\ref{lim_wf_0e}),
transforms for increasing disorder to an oscillatory line in the
$\kappa$-direction at $x=j+N/2$.  

\begin{figure}[ht]
\centerline{\includegraphics[clip=,width=0.99\columnwidth]{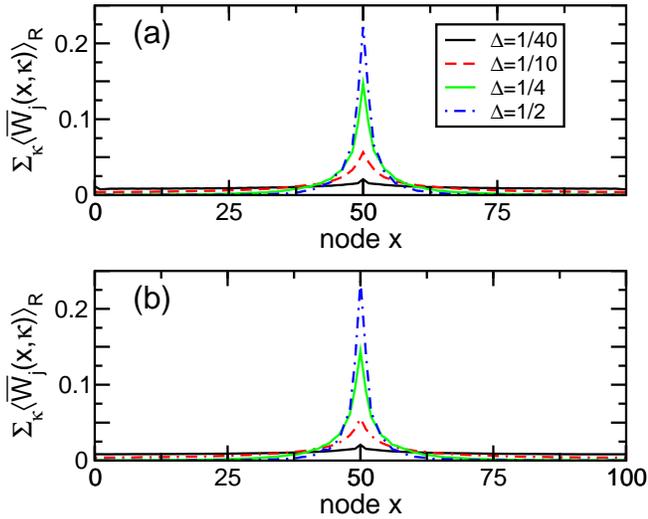}}
\caption{(Color online)
DOD: Marginal distribution $\sum_\kappa \langle{\overline
W}_j(x,\kappa)\rangle_R$ for different $\Delta$ and (a) $N=100$ and (b)
$N=101$.}
\label{marg_lim_wigner}
\end{figure}

By summing now again along the $\kappa$-direction, this peculiar behavior
vanishes and also for even $N$ we get a localized marginal limiting
probability distribution at $x=j$.  Figure \ref{marg_lim_wigner} shows the
marginal distribution $\sum_\kappa \langle{\overline W}_j(x,\kappa)\rangle_R$ for even
and odd $N$ with DOD. As expected, for even $N$,
Fig.~\ref{marg_lim_wigner}(a), there are no remainders of the peculiar
distribution of $\langle{\overline W}_j(x,\kappa)\rangle_R$ at $x=j+N/2$
for large $\Delta$. Furthermore, for high disorder, the marginal
distributions of
$\langle{\overline W}_j(x,\kappa)\rangle_R$ have a shape similar to that
of $\langle W_j(x,\kappa,t)\rangle_R$, compare
Figs.~\ref{wigner_marg}(a) and \ref{marg_lim_wigner}(b).

In general, the difference in the long time average $\sum_\kappa
\langle{\overline W}_j(x,\kappa)\rangle_R$ between even and odd $N$
strongly depends on the disorder. Without disorder, the two cases are
distinct \cite{mb2006a}. For odd $N$ the constructive interference at the
node $j$ leads to only one peak in $\sum_\kappa \langle{\overline
W}_j(x,\kappa)\rangle_R$, see Eq.~(20) in \cite{mb2006a}. However, for
even $N$ the constructive interference at nodes $j$ and also at node
$j+N/2$ leads to two peaks in $\sum_\kappa \langle{\overline
W}_j(x,\kappa)\rangle_R$, see Eq.~(21) in \cite{mb2006a}. With increasing
disorder, the second peak in $\sum_\kappa \langle{\overline
W}_j(x,\kappa)\rangle_R$ for even $N$ at $x=j+N/2$ vanishes and gives rise
to only one peak at $x=j$, see Fig.~\ref{marg_lim_wigner}. Nonetheless,
for large but finite $N$ the WF itself [given in
Eq.(\ref{wigner_discrete})] still allows us to distinguish between the odd
and the even case, see Figs.~\ref{lim_wigner} and \ref{lim_wigner_even}.

\section{Conclusions}

We have analyzed the effect of static disorder on the coherent exciton
transport by means of discrete Wigner functions.  We have studied
numerically the dynamics on a ring of $N$ sites in the presence of pure
diagonal disorder and also of diagonal and off-diagonal disorder.  The
previously found characteristic patterns of the unperturbed WF in the
quantum mechanical phase space are destroyed by the disorder. Instead, the
WF shows strong localization about the initial node. Integrating out the
details of the time evolution by considering the long time average of the
WF, shows an even more pronounced localization.

\section*{Acknowledgments}

This work was supported by a grant from the Ministry of Science, Research
and the Arts of Baden-W\"urttemberg (AZ: 24-7532.23-11-11/1). Further
support from the Deutsche Forschungsgemeinschaft (DFG) and the Fonds der
Chemischen Industrie is gratefully acknowledged.


\begin{thebibliography}{28}
\expandafter\ifx\csname natexlab\endcsname\relax\def\natexlab#1{#1}\fi
\expandafter\ifx\csname bibnamefont\endcsname\relax
  \def\bibnamefont#1{#1}\fi
\expandafter\ifx\csname bibfnamefont\endcsname\relax
  \def\bibfnamefont#1{#1}\fi
\expandafter\ifx\csname citenamefont\endcsname\relax
  \def\citenamefont#1{#1}\fi
\expandafter\ifx\csname url\endcsname\relax
  \def\url#1{\texttt{#1}}\fi
\expandafter\ifx\csname urlprefix\endcsname\relax\def\urlprefix{URL }\fi
\providecommand{\bibinfo}[2]{#2}
\providecommand{\eprint}[2][]{\url{#2}}

\bibitem[{\citenamefont{Wigner}(1932)}]{Wigner1932}
\bibinfo{author}{\bibfnamefont{E.~P.} \bibnamefont{Wigner}},
  \bibinfo{journal}{Phys.\ Rev.} \textbf{\bibinfo{volume}{40}},
  \bibinfo{pages}{749} (\bibinfo{year}{1932}).

\bibitem[{\citenamefont{Hillery et~al.}(1984)\citenamefont{Hillery, O'Connell,
  Scully, and Wigner}}]{Hillery1984}
\bibinfo{author}{\bibfnamefont{M.}~\bibnamefont{Hillery}},
  \bibinfo{author}{\bibfnamefont{R.~F.} \bibnamefont{O'Connell}},
  \bibinfo{author}{\bibfnamefont{M.~O.} \bibnamefont{Scully}},
  \bibnamefont{and} \bibinfo{author}{\bibfnamefont{E.~P.}
  \bibnamefont{Wigner}}, \bibinfo{journal}{Phys.\ Rep.}
  \textbf{\bibinfo{volume}{106}}, \bibinfo{pages}{121} (\bibinfo{year}{1984}).

\bibitem[{\citenamefont{Schleich}(2001)}]{Schleich}
\bibinfo{author}{\bibfnamefont{W.~P.} \bibnamefont{Schleich}},
  \emph{\bibinfo{title}{Quantum Optics in Phase Space}}
  (\bibinfo{publisher}{Wiley-VCH, Berlin}, \bibinfo{year}{2001}).

\bibitem[{\citenamefont{Mandel and Wolf}(1995)}]{Mandel-Wolf}
\bibinfo{author}{\bibfnamefont{L.}~\bibnamefont{Mandel}} \bibnamefont{and}
  \bibinfo{author}{\bibfnamefont{E.}~\bibnamefont{Wolf}},
  \emph{\bibinfo{title}{Optical Coherence and Quantum Optics}}
  (\bibinfo{publisher}{Cambridge University Press, Cambridge, England},
  \bibinfo{year}{1995}).

\bibitem[{\citenamefont{Kluksdahl et~al.}(1989)\citenamefont{Kluksdahl, Kriman,
  Ferry, and Ringhofer}}]{kluksdahl1989}
\bibinfo{author}{\bibfnamefont{N.~C.} \bibnamefont{Kluksdahl}},
  \bibinfo{author}{\bibfnamefont{A.~M.} \bibnamefont{Kriman}},
  \bibinfo{author}{\bibfnamefont{D.~K.} \bibnamefont{Ferry}}, \bibnamefont{and}
  \bibinfo{author}{\bibfnamefont{C.}~\bibnamefont{Ringhofer}},
  \bibinfo{journal}{Phys.\ Rev.\ B} \textbf{\bibinfo{volume}{39}},
  \bibinfo{pages}{7720} (\bibinfo{year}{1989}).

\bibitem[{\citenamefont{Buot}(1993)}]{buot1993}
\bibinfo{author}{\bibfnamefont{F.~A.} \bibnamefont{Buot}},
  \bibinfo{journal}{Phys.\ Rep.} \textbf{\bibinfo{volume}{234}},
  \bibinfo{pages}{73} (\bibinfo{year}{1993}).

\bibitem[{\citenamefont{Bordone et~al.}(1999)\citenamefont{Bordone, Pascoli,
  Brunetti, Bertoni, Jacoboni, and Abramo}}]{bordone1999}
\bibinfo{author}{\bibfnamefont{P.}~\bibnamefont{Bordone}},
  \bibinfo{author}{\bibfnamefont{M.}~\bibnamefont{Pascoli}},
  \bibinfo{author}{\bibfnamefont{R.}~\bibnamefont{Brunetti}},
  \bibinfo{author}{\bibfnamefont{A.}~\bibnamefont{Bertoni}},
  \bibinfo{author}{\bibfnamefont{C.}~\bibnamefont{Jacoboni}}, \bibnamefont{and}
  \bibinfo{author}{\bibfnamefont{A.}~\bibnamefont{Abramo}},
  \bibinfo{journal}{Phys.\ Rev.\ B} \textbf{\bibinfo{volume}{59}},
  \bibinfo{pages}{3060} (\bibinfo{year}{1999}).

\bibitem[{\citenamefont{Anderson}(1958)}]{anderson1958}
\bibinfo{author}{\bibfnamefont{P.~W.} \bibnamefont{Anderson}},
  \bibinfo{journal}{Phys.\ Rev.} \textbf{\bibinfo{volume}{109}},
  \bibinfo{pages}{1492} (\bibinfo{year}{1958}).

\bibitem[{\citenamefont{Abrahams et~al.}(1979)\citenamefont{Abrahams, Anderson,
  Licciardello, and Ramakrishnan}}]{abrahams1979}
\bibinfo{author}{\bibfnamefont{E.}~\bibnamefont{Abrahams}},
  \bibinfo{author}{\bibfnamefont{P.~W.} \bibnamefont{Anderson}},
  \bibinfo{author}{\bibfnamefont{D.~C.} \bibnamefont{Licciardello}},
  \bibnamefont{and} \bibinfo{author}{\bibfnamefont{T.~V.}
  \bibnamefont{Ramakrishnan}}, \bibinfo{journal}{Phys.\ Rev.\ Lett.}
  \textbf{\bibinfo{volume}{42}}, \bibinfo{pages}{673} (\bibinfo{year}{1979}).

\bibitem[{\citenamefont{Anderson}(1978)}]{anderson1978}
\bibinfo{author}{\bibfnamefont{P.~W.} \bibnamefont{Anderson}},
  \bibinfo{journal}{Rev.\ Mod.\ Phys.} \textbf{\bibinfo{volume}{50}},
  \bibinfo{pages}{191} (\bibinfo{year}{1978}).

\bibitem[{\citenamefont{Logan and Wolynes}(1987)}]{logan1987}
  \bibinfo{author}{\bibfnamefont{D.~E.} \bibnamefont{Logan}}
  \bibnamefont{and} \bibinfo{author}{\bibfnamefont{P.~G.}
  \bibnamefont{Wolynes}},
   \bibinfo{journal}{J.\ Chem.\ Phys.} \textbf{\bibinfo{volume}{87}},
   \bibinfo{pages}{7199} (\bibinfo{year}{1987}).
		      
\bibitem[{\citenamefont{Abramavicius et~al.}(2004)\citenamefont{Abramavicius, Valkunas, and van Grondelle}}]{abramavicius2004}
  \bibinfo{author}{\bibfnamefont{D.}~\bibnamefont{Abramavicius}},
  \bibinfo{author}{\bibfnamefont{L.}~\bibnamefont{Valkunas}},
  \bibnamefont{and} \bibinfo{author}{\bibfnamefont{R.}~\bibnamefont{van
  Grondelle}},
  \bibinfo{journal}{Phys.\ Chem.\ Chem.\ Phys.}
  \textbf{\bibinfo{volume}{6}}, \bibinfo{pages}{3097}
  (\bibinfo{year}{2004}).

 \bibitem[{\citenamefont{Dominguez-Adame and Malyshev}(2004)}]{dominguez2004}
  \bibinfo{author}{\bibfnamefont{F.}~\bibnamefont{Dominguez-Adame}}
  \bibnamefont{and} \bibinfo{author}{\bibfnamefont{V.~A.}
  \bibnamefont{Malyshev}}, 
  \bibinfo{journal}{Am.\ J.\ Phys.} \textbf{\bibinfo{volume}{72}},
  \bibinfo{pages}{226} (\bibinfo{year}{2004}).

\bibitem[{\citenamefont{Helmes et~al.}(2005)\citenamefont{Helmes, Sindel, Borda, and von Delft}}]{helmes2005}
  \bibinfo{author}{\bibfnamefont{R.~W.} \bibnamefont{Helmes}},
  \bibinfo{author}{\bibfnamefont{M.}~\bibnamefont{Sindel}},
  \bibinfo{author}{\bibfnamefont{L.}~\bibnamefont{Borda}},
  \bibnamefont{and} \bibinfo{author}{\bibfnamefont{J.}~\bibnamefont{von
  Delft}},
  \bibinfo{journal}{Phys.\ Rev.\ B} \textbf{\bibinfo{volume}{72}},
  \bibinfo{pages}{125301} (\bibinfo{year}{2005}).
		        
\bibitem[{\citenamefont{Barford and Duffy}(2006)}]{barford2006}
  \bibinfo{author}{\bibfnamefont{W.}~\bibnamefont{Barford}}
  \bibnamefont{and} \bibinfo{author}{\bibfnamefont{C.~D.~P.}
  \bibnamefont{Duffy}},
  \bibinfo{journal}{Phys.\ Rev.\ B} \textbf{\bibinfo{volume}{74}},
  \bibinfo{pages}{075207} (\bibinfo{year}{2006}).
			      
			    
\bibitem[{\citenamefont{Keating et~al.}(2006)\citenamefont{Keating,
	Linden, Matthews, and Winter}}]{keating2006}
  \bibinfo{author}{\bibfnamefont{J.~P.} \bibnamefont{Keating}},
  \bibinfo{author}{\bibfnamefont{N.}~\bibnamefont{Linden}},
  \bibinfo{author}{\bibfnamefont{J.~C.~F.} \bibnamefont{Matthews}},
  \bibnamefont{and}
  \bibinfo{author}{\bibfnamefont{A.}~\bibnamefont{Winter}},
  \bibinfo{journal}{arXiv:} \bibinfo{pages}{quant--ph/0606205}
  (\bibinfo{year}{2006}). 

\bibitem[{\citenamefont{Farhi and Gutmann}(1998)}]{farhi1998}
\bibinfo{author}{\bibfnamefont{E.}~\bibnamefont{Farhi}} \bibnamefont{and}
  \bibinfo{author}{\bibfnamefont{S.}~\bibnamefont{Gutmann}},
  \bibinfo{journal}{Phys.\ Rev.\ A} \textbf{\bibinfo{volume}{58}},
  \bibinfo{pages}{915} (\bibinfo{year}{1998}).

\bibitem[{\citenamefont{M{\"u}lken et~al.}(2006)\citenamefont{M{\"u}lken,
	Bierbaum, and Blumen}}]{mbb2006a}
  \bibinfo{author}{\bibfnamefont{O.}~\bibnamefont{M{\"u}lken}},
  \bibinfo{author}{\bibfnamefont{V.}~\bibnamefont{Bierbaum}},
  \bibnamefont{and}
  \bibinfo{author}{\bibfnamefont{A.}~\bibnamefont{Blumen}},
  \bibinfo{journal}{J.\ Chem.\ Phys.} \textbf{\bibinfo{volume}{124}},
  \bibinfo{pages}{124905} (\bibinfo{year}{2006}).

\bibitem[{\citenamefont{Kenkre and Reineker}(1982)}]{Kenkre}
\bibinfo{author}{\bibfnamefont{V.~M.} \bibnamefont{Kenkre}} \bibnamefont{and}
  \bibinfo{author}{\bibfnamefont{P.}~\bibnamefont{Reineker}},
  \emph{\bibinfo{title}{Exciton Dynamics in Molecular Crystals and Aggregates}}
  (\bibinfo{publisher}{Springer, Berlin}, \bibinfo{year}{1982}).

\bibitem[{\citenamefont{Shore}(1990)}]{Shore}
\bibinfo{author}{\bibfnamefont{B.~W.} \bibnamefont{Shore}},
  \emph{\bibinfo{title}{The Theory of Coherent Atomic Excitation}}
  (\bibinfo{publisher}{Wiley, New York}, \bibinfo{year}{1990}).

\bibitem[{\citenamefont{Haken}(1976)}]{Haken}
\bibinfo{author}{\bibfnamefont{H.}~\bibnamefont{Haken}},
  \emph{\bibinfo{title}{Quantum Field Theory of Solids}}
  (\bibinfo{publisher}{North-Holland, Amsterdam}, \bibinfo{year}{1976}).

\bibitem[{\citenamefont{M{\"u}lken and Blumen}(2006)}]{mb2006a}
\bibinfo{author}{\bibfnamefont{O.}~\bibnamefont{M{\"u}lken}} \bibnamefont{and}
  \bibinfo{author}{\bibfnamefont{A.}~\bibnamefont{Blumen}},
  \bibinfo{journal}{Phys.\ Rev.\ A} \textbf{\bibinfo{volume}{73}},
  \bibinfo{pages}{012105} (\bibinfo{year}{2006}).

\bibitem[{\citenamefont{M{\"u}lken and Blumen}(2005)}]{mb2005b}
\bibinfo{author}{\bibfnamefont{O.}~\bibnamefont{M{\"u}lken}} \bibnamefont{and}
  \bibinfo{author}{\bibfnamefont{A.}~\bibnamefont{Blumen}},
  \bibinfo{journal}{Phys.\ Rev.\ E} \textbf{\bibinfo{volume}{71}},
  \bibinfo{pages}{036128} (\bibinfo{year}{2005}).

\bibitem[{\citenamefont{Wootters}(1987)}]{Wootters1987}
\bibinfo{author}{\bibfnamefont{W.~K.} \bibnamefont{Wootters}},
  \bibinfo{journal}{Ann.\ Phys.} \textbf{\bibinfo{volume}{176}},
  \bibinfo{pages}{1} (\bibinfo{year}{1987}).

\bibitem[{\citenamefont{Cohendet et~al.}(1988)\citenamefont{Cohendet, Combe,
  Sirugeu, and Sirugue-Collin}}]{Cohendet1988}
\bibinfo{author}{\bibfnamefont{O.}~\bibnamefont{Cohendet}},
  \bibinfo{author}{\bibfnamefont{P.}~\bibnamefont{Combe}},
  \bibinfo{author}{\bibfnamefont{M.}~\bibnamefont{Sirugeu}}, \bibnamefont{and}
  \bibinfo{author}{\bibfnamefont{M.}~\bibnamefont{Sirugue-Collin}},
  \bibinfo{journal}{J.\ Phys.\ A} \textbf{\bibinfo{volume}{21}},
  \bibinfo{pages}{2875} (\bibinfo{year}{1988}).

\bibitem[{\citenamefont{Leonhardt}(1995)}]{Leonhardt1995}
\bibinfo{author}{\bibfnamefont{U.}~\bibnamefont{Leonhardt}},
  \bibinfo{journal}{Phys.\ Rev.\ Lett.} \textbf{\bibinfo{volume}{74}},
  \bibinfo{pages}{4101} (\bibinfo{year}{1995}).

\bibitem[{\citenamefont{Takami et~al.}(2001)\citenamefont{Takami, Hashimoto,
  Horibe, and Hayashi}}]{Takami2001}
\bibinfo{author}{\bibfnamefont{A.}~\bibnamefont{Takami}},
  \bibinfo{author}{\bibfnamefont{T.}~\bibnamefont{Hashimoto}},
  \bibinfo{author}{\bibfnamefont{M.}~\bibnamefont{Horibe}}, \bibnamefont{and}
  \bibinfo{author}{\bibfnamefont{A.}~\bibnamefont{Hayashi}},
  \bibinfo{journal}{Phys.\ Rev.\ A} \textbf{\bibinfo{volume}{64}},
  \bibinfo{pages}{032114} (\bibinfo{year}{2001}).

\bibitem[{\citenamefont{Miquel et~al.}(2002)\citenamefont{Miquel, Paz, and
  Saraceno}}]{Miquel2002}
\bibinfo{author}{\bibfnamefont{C.}~\bibnamefont{Miquel}},
  \bibinfo{author}{\bibfnamefont{J.~P.} \bibnamefont{Paz}}, \bibnamefont{and}
  \bibinfo{author}{\bibfnamefont{M.}~\bibnamefont{Saraceno}},
  \bibinfo{journal}{Phys.\ Rev.\ A} \textbf{\bibinfo{volume}{65}},
  \bibinfo{pages}{062309} (\bibinfo{year}{2002}).

\end{thebibliography}
\end{document}